\documentclass[showpacs,prl,twocolumn,superscriptaddress]{revtex4}
\usepackage{graphics,graphicx,amssymb,amsmath}
\newcommand{\etal}{{\it et al.}}

\begin{document}

\title{Correlated impurities and intrinsic spin liquid physics in the kagome material Herbertsmithite}

\author{Tian-Heng Han$^{\dag}$}
\affiliation{Materials Science Division, Argonne National Laboratory, Argonne, IL 60439}
\affiliation{James Franck Institute and Department of Physics, University of Chicago, Chicago, IL  60637}
\author{M. R. Norman}
\affiliation{Materials Science Division, Argonne National Laboratory, Argonne, IL 60439}
\author{J.-J. Wen}
\affiliation{Department of Applied Physics, Stanford University, Stanford, CA  94305}
\affiliation{Stanford Institute for Materials and Energy Sciences, SLAC National Accelerator Laboratory, 2575 Sand Hill Road, Menlo Park, CA 94025}
\author{Jose A. Rodriguez-Rivera}
\affiliation{NIST Center for Neutron Research, National Institute of Standards and Technology, Gaithersburg, Maryland  20899}
\affiliation{Department of Materials Science, University of Maryland, College Park, Maryland  20742}
\author{Joel S. Helton}
\affiliation{Department of Physics, The United States Naval Academy, Annapolis, MD  21402}
\affiliation{NIST Center for Neutron Research, National Institute of Standards and Technology, Gaithersburg, Maryland  20899}
\author{Collin Broholm}
\affiliation{Institute for Quantum Matter and Department of Physics and Astronomy, The Johns Hopkins University, Baltimore, MD  21218}
\affiliation{NIST Center for Neutron Research, National Institute of Standards and Technology, Gaithersburg, Maryland  20899}
\author{Young S. Lee$^{\dag}$}
\affiliation{Department of Applied Physics, Stanford University, Stanford, CA  94305}
\affiliation{Stanford Institute for Materials and Energy Sciences, SLAC National Accelerator Laboratory, 2575 Sand Hill Road, Menlo Park, CA 94025}

\begin{abstract}
Low energy inelastic neutron scattering on single crystals of the kagome spin liquid compound ZnCu$_3$(OD)$_6$Cl$_2$
(Herbertsmithite) reveals antiferromagnetic correlations between impurity spins for energy transfers $\hbar \omega < 0.8$~meV ($\sim$$J/20$).
The momentum dependence differs significantly from higher energy scattering which arises from the intrinsic kagome spins. The low energy fluctuations are characterized by diffuse scattering near wavevectors (100) and (00$\frac{3}{2}$), which is consistent with antiferromagnetic correlations between pairs of nearest neighbor Cu impurities on adjacent triangular (Zn) interlayers. The corresponding impurity lattice resembles a simple cubic lattice in the dilute limit below the percolation threshold. Such an impurity model can describe prior neutron, NMR, and specific heat data. The low energy neutron data are consistent with the presence of a small spin-gap ($\Delta \sim$~0.7 meV) in the kagome layers, similar to that recently observed by NMR. The ability to distinguish the scattering due to Cu impurities from that of the planar kagome Cu spins provides a new avenue for probing intrinsic spin liquid physics. \end{abstract}

\date{\today}

\maketitle

The synthesis of ZnCu$_3$(OH)$_6$Cl$_2$ \cite{shores} (Herbertsmithite) was a watershed in the field
of frustrated magnetism.  The spin 1/2 states on the Cu sites, which sit on a kagome lattice, are characterized 
by a sizable superexchange interaction of about 200 K, but the spins do not order or freeze down to the lowest temperatures measured,
suggesting that the ground state is a quantum spin liquid \cite{helton1,mendels1}. However, in order to properly classify the ground state, a more detailed understanding of the low energy properties is required. Leading theoretical work, such as density matrix renormalization
group calculations of the Heisenberg model for S=1/2 spins on a kagome lattice \cite{white,depenbrock,jiang}, indicates the ground state is a topological spin liquid with a  spin gap that is small relative to the exchange energy $J$.

One challenge in determining the intrinsic low energy properties is a small fraction of Cu impurities in the samples. The bulk spin susceptibility exhibits a diverging Curie-like tail, an indication that some of the Cu spins act like weakly coupled impurities. This is also consistent with low temperature specific heat data \cite{helton1}. The nature of these defects has been a substantial controversy in the field \cite{mendels1,mendels2}.  However, the advent of high quality single crystals has led to a qualitative improvement of our understanding of Herbertsmithite \cite{han1}. Single crystal NMR \cite{imai2} and resonant x-ray diffraction \cite{freedman} data have shown that the impurities are 15\% Cu on triangular Zn intersites while the kagome planes are fully occupied with Cu. Moreover, recent NMR data indicate the kagome spins possess a spin gap of $\sim$0.9 meV \cite{imai3}, implying that the divergent response seen in the dynamic spin susceptibility below 1 meV \cite{helton2} is due to impurities. Since neutron scattering is a bulk probe of all of the spins, low energy scattering from impurities may obscure the intrinsic response of the kagome spins below the gap energy \cite{han2}.

\begin{figure*}
\includegraphics[width=16cm]{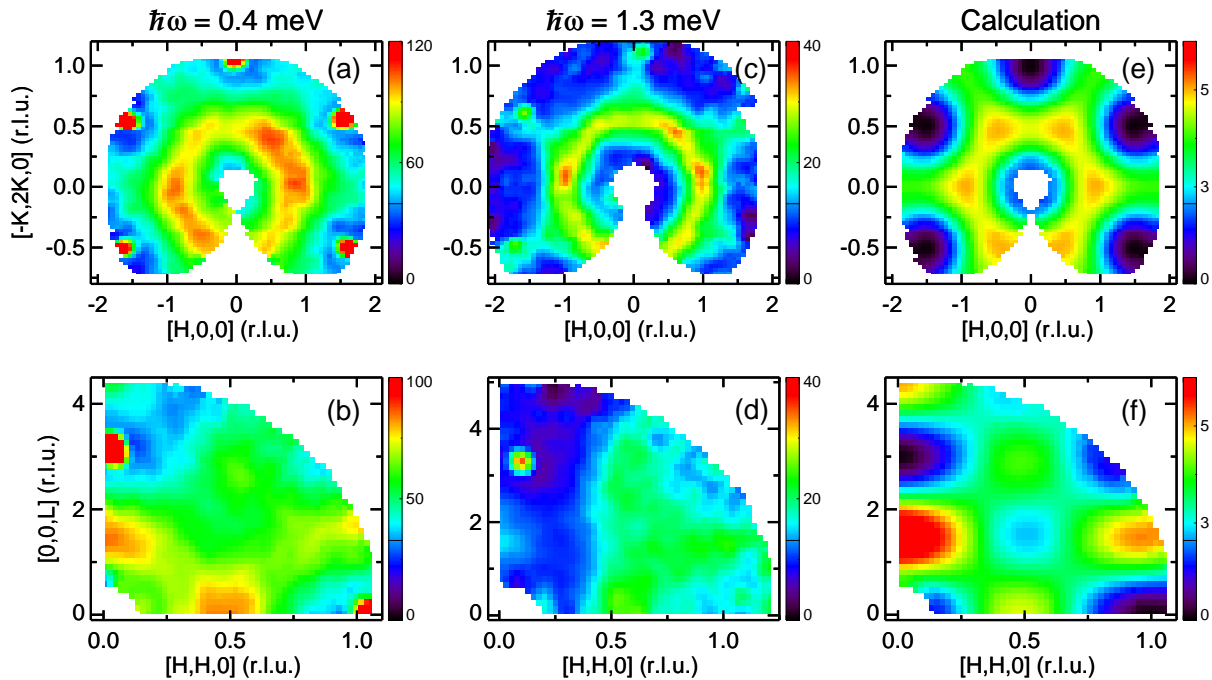}
\caption{(color online) (a)-(d) Inelastic neutron data on Herbertsmithite in the (HK0) and (HHL) scattering planes at $T=2$K for $\hbar \omega = 0.4$ meV and $\hbar \omega = 1.3$ meV. The bright spots at (110) and (003) arise from structural Bragg peaks. The diffuse spots at (100), (00$\frac{3}{2}$) and
($\frac{1}{2}$$\frac{1}{2}$0) are magnetic in origin. Note that the (00$\frac{3}{2}$) diffuse spot is particularly pronounced at 0.4 meV, while the magnetic scattering at 1.3 meV is nearly independent of $L$. (e)-(f) Plots of the calculated ${\cal S}({\bf Q})$ in the (HK0) and the (HHL) planes, representing antiferromagnetically correlated nearest neighbor impurities on the interlayer sites, as described in the text.
}
\label{fig1}
\end{figure*}

Here, we report high resolution inelastic neutron scattering measurements on single crystals which allow us to distinguish the scattering from impurity spins from that of the intrinsic kagome layer spins. Experiments were performed using the upgraded Multi-Axis Crystal Spectrometer (MACS) at  NIST. A pumped helium cryostat  cooled the sample to $T=2$~K.  The final analyzed neutron energy was $E_{f}=3.7$~meV with an energy resolution of 0.15~meV (full-width at half-maximum). Cooled polycrystalline Be and BeO filters were in place before and after the sample, respectively. Single crystals of deuterated Herbertsmithite were prepared as previously reported \cite{han2}.  Fifteen of the largest pieces were co-aligned on an aluminum sample holder, yielding a total mass of 1.2 grams with an overall sample mosaic of $\sim$~2$^{\circ}$.  The background was measured with the empty sample holder inside the cryostat for every instrumental configuration used and subtracted from the corresponding sample measurements.

Prior inelastic neutron scattering measurements on single crystals by some of us \cite{han2} revealed a continuum of scattering consistent with fractionalized spinon excitations. That study primarily focused on energy transfers from about 0.75 meV to 11 meV \cite{han2}.  The response in the ($HK0$) plane above 1 meV forms a continuum, consistent with a singlet form factor involving nearest neighbor kagome spins. Below this energy, though, the momentum pattern was found to feature broad spots with maxima at (100) and equivalent positions. Here, we have acquired new data in the ($HK0$) scattering plane at $\hbar \omega = 0.4$ meV and 1.3 meV, as shown in Fig.~1(a) and (c).  The ${\bf Q}$-dependence of the scattering at fixed energy transfer shows a distinct rotation relative to the high energy dimer-like pattern with maxima near ($\frac{2}{3}\frac{2}{3}$0) to a low energy pattern with maxima at (100). One can imagine various ways in which enhanced scattering at (100) might emerge: for example, kagome spins with dynamical $q=0$ correlations (as observed in iron jarosite KFe$_3$(OH)$_6$(SO$_4$)$_2$ \cite{grohol}) as well as a ferromagnetic arrangement of impurity spins within the interlayers could give rise to such peaks. However, it may be necessary to go beyond 2D models, since the interaction pathways between the interlayer Cu impurities would imply correlations along the $c$-direction as well.

Therefore, we have performed additional measurements in the ($HHL$) scattering plane which allow us to probe both intralayer and interlayer correlations. These measurements reveal that the lowest energy fluctuations have short range correlations along all three crystallographic directions. As shown in Fig.~1(b), diffuse peaks  are seen at the (00$\frac{3}{2}$) and ($\frac{1}{2}$$\frac{1}{2}$0) positions for $\hbar \omega = 0.4$~meV. This intensity emerges below an energy scale of $\sim$0.8 meV where an enhanced dynamic magnetic response was previously reported \cite{helton2,han2}. The diffuse peak at $L=\frac{3}{2}$ has the same position along $L$ as the magnetic Bragg peaks  in iron jarosite \cite{jaro,matan} where long-range order yields a magnetic cell that is doubled along the $c$-axis \cite{foot1}. In contrast, the scattering at a higher energy of  $\hbar \omega = 1.3$~meV (Fig.~1d) shows little variation along the $L$-direction, consistent with quasi-two-dimensional correlations as expected for intrinsic kagome spins. This new observation establishes a clear dichotomy between the low energy 3D excitations (below 0.8 meV) and the higher energy 2D excitations. The explicit observation of quasi-2D correlations confirms that the spin excitations measured above 1 meV by Han {\it et al.}~\cite{han2} essentially derive from the two-dimensional physics of a single kagome lattice. Moreover, the dichotomy implies that the physics at low energies (such as effects of weakly coupled impurities) quickly diminishes at the higher measured energies. Hence, it appears neutron scattering can distinguish the intrinsic response of  kagome spins from interlayer impurity spin correlations through their distinct $\bf Q$-dependences. 

\begin{figure}
\includegraphics[width=8.7cm]{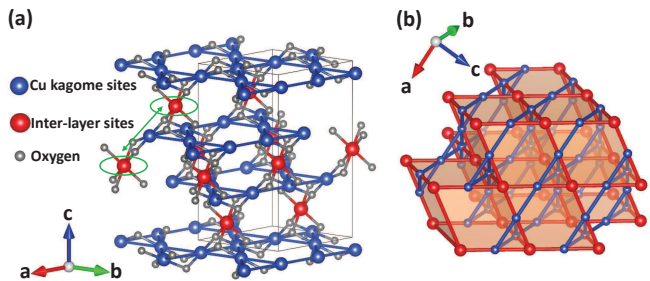}
\caption{(color online) (a) Proposed model of the antiferromagnetically correlated nearest neighbor Cu impurities which give rise to the low energy scattering. The blue spheres are the intrinsic kagome Cu sites. The red spheres denote the Zn interlayer sites. The green ovals indicate interlayer sites which have Cu impurities on nearest neighbor locations. These moments are fluctuating with AF correlations. In our sample, only about 15\% of the Zn sites have a Cu impurity. (b) The effective connectivity of the impurity lattice as a simple cubic lattice. Here, the red bonds indicate the NN impurity-impurity bonds identified in the structure factor calculations. The blue spheres and bonds indicate the intrinsic kagome layers.
}
\label{fig2}
\end{figure}

In our sample,  $x=15$\% of the Zn sites are occupied by Cu. If these are randomly distributed, then $1-(1-x)^6=62$\% of all impurities have at least one other impurity on a nearest neighbor (NN) site that resides on an adjacent interlayer plane (Fig.~2a). The wave vector dependence of scattering from short range correlations involving impurity spins can be modeled by a correlation function of the form
 ${\cal S}({\bf Q})=N |F(Q)|^2 \langle S^2 \rangle (1 + \sum_n \frac{2 m_n}{N} f_n({\bf Q}) \frac{\langle SS^\prime\rangle_{n}}{\langle S^2 \rangle})$, where the sum is over bonds between the impurity to its $n$th nearest neighbor. Here, $N$ is the number of impurity spins, $F(Q)$ is the Cu form factor \cite{formfac}, $m_n$ is the number of bonds for a given $n$, and $f_n({\bf Q})$ denotes the sum of $\cos{({\bf Q} \cdot {\bf r}_n)}$ over all bond directions (divided by the number of bond directions). In this notation, $f_1$ is the correlator between an impurity spin and its nearest neighbor kagome spins, and $f_3$ is between an impurity spin and its nearest neighbor impurity spins (indicated by the green ovals in Fig.~2). Note that $f_2$ (between an impurity spin and the next nearest neighbor kagome spin) and $f_n$ for $n>3$ denote bonds with exchange pathways that are unlikely to have significant strength.

Indeed, we find that $f_3$ dominates the correlation function. In Fig.~1(e) and (f), we show that a model including only $n=3$ pairs (impurity-impurity pairs) with antiferromagnetic (AF) correlations ($\langle SS^\prime\rangle_3<0$) provides a good description of the data. The low energy diffuse intensities in both the ($HK0$) and ($HHL$) zones are well described. Fig.~3 shows magnetic diffuse scattering along $(00L)$-direction obtained by subtracting $T=20$ K data from $T=2$ K data. We see this is well fit by an AF $n=3$ term only (blue dashed line) with no evidence for correlations to kagome spins which would give a different periodicity along $L$ (adding a component of $f_1$ does not improve the fit, as shown by the solid red line). Such an analysis places a limit on correlations to the kagome layer: $|\langle SS^\prime\rangle_{1}/ \langle SS^\prime\rangle_{3}|<0.01$. Correlations between impurity spins to further neighbor impurity spins within the same triangular lattice plane can similarly be neglected. This establishes the impurity spin network as based on AF interlayer impurity interactions. The corresponding lattice has the connectivity of a simple cubic lattice (Fig.~2(b)) with a percolation threshold of 0.3116 \cite{PhysRevE.87.052107}. Hence, the impurities in Herbertsmithite are in the dilute limit.

\begin{figure}
\includegraphics[width=7.0cm]{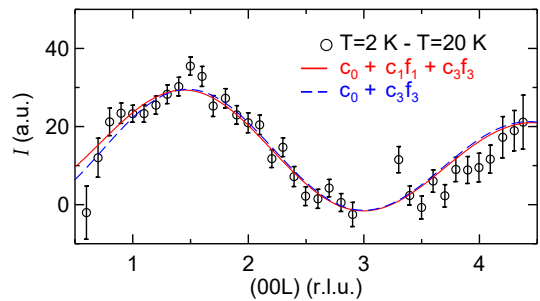}
\caption{(color online) Line scan through the data obtained by subtracting the data along (00$L$) at 0.4 meV acquired at $T=20$K from the data at $T=2$K.  Here, to improve statistics, an integration width of $\pm$0.15 was used along the ($HH0$) direction.  The blue dashed line denotes a fit to these data involving only the correlator between nearest neighbor impurities $f_3$. The red solid line denote a fit that also includes the correlator between the impurity and the nearest spins on the kagome layers $f_1$, however the contribution of $f_1$ is found to be small. Error bars indicate 1$\sigma$.
}
\label{fig3}
\end{figure}

A possible motivation for AF correlations between NN impurities may be derived from the magnetic structure of the pure Cu analogue, Cu$_2$(OD)$_3$Cl (clinoatacamite) \cite{kim,wills}.  Clinoatacamite has a monoclinically distorted structure, with two different kagome sites, consistent with a Jahn-Teller distortion associated with the Cu intersites \cite{janson}. Here, a given interlayer up spin is connected to two down spins and one up spin on the neighboring interlayer (above or below).  This indicates a net AF coupling between nearest neighbor intersites \cite{foot2}. 

While 62\% of the Cu impurities have at least one nearest neighbor impurity on an adjacent triangular plane, only $x^2=2$\% of the kagome Cu spins are located directly between pairs of impurities. The remaining 38\% of impurities that do not form an interlayer singlet are to first order decoupled from the impurity spin cluster. This indicates  the intrinsic kagome physics may not be significantly disturbed by the correlated impurities. However, low energy measurements that average over the bulk are likely dominated by the weakly coupled impurities on NN intersites, consistent with the small AF Curie-Weiss temperature ($\sim 1$ K) inferred from the low temperature susceptibility \cite{bert,han4}. If the presence of Cu ions on the Zn intersites causes a local distortion of the Herbertsmithite structure to resemble that of clinoatacamite, this may at least partially explain the Cl NMR data which indicates a local structural distortion below 150K \cite{imai1}. It is also consistent with the anisotropy of the low temperature bulk susceptibility, which indicates impurities occupy an anisotropic environment \cite{han4}.  In this context, we can also mention a recent study of Zn-doped paratacamite, where two different interlayer sites was identified \cite{welch}.

\begin{figure}[t!]
\includegraphics[width=8.1cm]{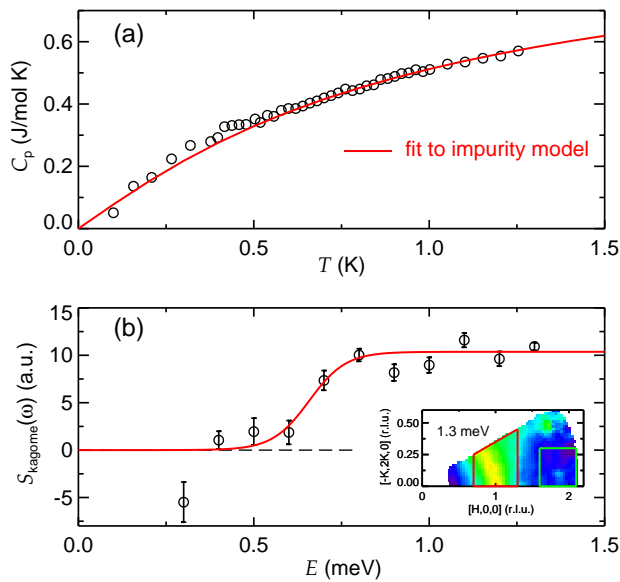} \vspace{0mm}
\caption{(color online) (a) Specific heat versus temperature in Herbertsmithite \cite{helton1}.  The solid curve denotes the equation described in the text based on the impurity model, with an assumed impurity concentration of 12\% (these data were taken on a powder sample, different from the single crystals used for the neutron measurements). (b) A measure of the intrinsic scattering ${\cal S}_{kagome}(\omega)$ obtained after subtracting the impurity contribution as described in the text. These data are integrated over a large region in reciprocal space. The red line is fit to $\tanh{(\frac{\omega-\Delta}{\Gamma})}$ with $\Delta=0.66$ and $\Gamma=0.1$ used as a guide to the eye. Inset: a reciprocal space map of the integration regions used to obtain ${\cal S}_{imp}(\omega)$ (green box near (200)) and ${\cal S}_{kag+imp}(\omega)$ (red box near (100)).
}
\label{fig4}
\end{figure}

To further test this model for the impurities, we turn to previous low energy neutron and specific heat results. Neutron data \cite{nilsen} taken between 0.1 and 0.7 meV was also interpreted as due to Cu intersite spins, with a spectrum that can be approximately fit by a Lorentzian, with a relaxation rate $\Gamma=0.23$ meV \cite{dho}. We  use this  to estimate the specific heat due to the dynamical spin fluctuations of the defects.  Assuming $\chi$ is of the form $f(q)f(\omega)$ as for a continuum of spin excitations \cite{han2}, the free energy is $F = \int \frac{d\omega}{2\pi} \coth(\frac{\omega}{2T}) Im \ln(\chi^{-1}) $ which yields a specific heat $C = \int \frac{d\omega}{\pi} \frac{\omega}{T^2 \sinh^2(\frac{\omega}{2T})} (\frac{\omega}{2T} \coth(\frac{\omega}{2T}) - 1) \tan^{-1}(\frac{\omega}{\Gamma})$. This function (multiplied by $Rx$, where $R$ is the gas constant and $x$ the impurity concentration) is plotted in Fig.~4a in comparison to the specific heat data of Helton \etal~\cite{helton1}.  The only adjustable parameter was the impurity concentration $x$ of 12\% for the powder sample. This  reinforces the idea that the scaling behavior noted by Helton \etal~\cite{helton2} is due to impurity spins with a distribution of interactions as one might expect from weakly interacting non-percolating spin-1/2 clusters. A random bond Heisenberg model with a distribution of exchange interactions going as $J^{-\alpha}$ that is generated by renormalization group flow from an initial distribution of $J$'s \cite{gupta} can fit the dynamic susceptibility of Herbertsmithite quite well for $\alpha \sim 2/3$ \cite{helton2}. Here, the random bond Heisenberg model is directly applicable to the impurity lattice rather than the kagome planes. Calculations based on the random bond Heisenberg model on the kagome lattice \cite{kawamura} do not predict the recently observed spin gap in the kagome layers of Herbertsmithite \cite{imai3}.

A key remaining question is to understand the \emph{low energy} excitations of the \emph{intrinsic} kagome spins. The present study allows us to subtract the impurity scattering from the total scattering (integrated over select {\bf Q} regions) to obtain a measure of the intrinsic scattering. In the inset of Fig.~4(b), we show two regions in reciprocal space, near (200) and (100), over which we integrate the scattering to obtain ${\cal S}_{imp}(\omega)$ and ${\cal S}_{kag+imp}(\omega)$, respectively. ${\cal S}_{imp}(\omega)$ is obtained since the correlated impurities give substantial scattering at (200) equivalent positions which corresponds to positions of minimal scattering from the intrinsic kagome spins \cite{han2}. On the other hand, near (100), ${\cal S}_{kag+imp}(\omega)$ denotes a combination of the intrinsic and impurity scattering since both have substantial structure factors. We calculate ${\cal S}_{kagome}(\omega) = {\cal S}_{kag+imp}(\omega) - a \; {\cal S}_{imp}(\omega)$ where the scale factor $a \simeq 1.7$ is determined by the best match of the scattering amplitudes between 0.3 meV and 0.5 meV and is consistent with our correlated impurity structure factor (within 30\%). The resulting data for ${\cal S}_{kagome}(\omega)$ is plotted in Fig.4(b). The most remarkable feature is the sudden drop in intensity as $\omega$ decreases below 0.7 meV. This is similar to the magnitude of the spin gap deduced by NMR in zero field of 0.9(3) meV \cite{imai3}. This is the first indication of a triplet spin gap seen by inelastic neutron scattering.

In summary, we find that a correlated spin impurity picture gives a good account of the low energy neutron scattering data on Herbertsmithite ($\hbar \omega < 0.8$ meV), involving AF correlations between nearest neighbor interlayer impurities. Such a picture could be further tested by local structural studies to look for the expected distortion of the lattice about the Cu defects in the Zn planes due to the Jahn-Teller effect. This also provides crucial guidance for future measurements at ($HKL$) positions where impurity effects can be minimized to enhance sensitivity to the intrinsic physics of the 2D kagome planes, such as investigating the wave vector dependence of the low energy spin gap.

T.-H. H. acknowledges the support of the Grainger Fellowship provided by the Department of Physics and the support from Eric Isaacs through the Provost's Office, University of Chicago (neutron scattering studies and data analysis). The work at Stanford and SLAC was supported by the U.S. Department of Energy (DOE), Office of Science, Basic Energy Sciences, Materials Sciences and Engineering Division, under Contract No. DE-AC02-76SF00515 (neutron data analysis). The work at Argonne was supported by the Materials Sciences and Engineering Division, Basic Energy Sciences, Office of Science, US DOE (theory and data analysis). C.B. was supported by the US DOE, Office of Basic Energy Sciences, Division of Materials Sciences and Engineering under Grant DE-FG02-08ER46544. This work utilized facilities supported in part by the National Science Foundation under Agreement No.~DMR-1508249.

$^{\dag}$ tianheng@alum.mit.edu, youngsl@stanford.edu


\begin{thebibliography}{99}
\bibitem{foot1}
What we observe, though, is not a fluctuating version of the $q=0$ structure of the Fe jarosite, since for the latter, 00L peaks vanish since the sum of the in-plane components of the spins on a given kagome triangle is zero.
\bibitem{foot2}
In clinoatacamite, each interlayer spin is surrounded by four kagome spins of the same sign, and two of the opposite sign.
In our context, this would imply that an impurity spin would ferromagnetically polarize the surrounding kagome
spins, which is consistent with NMR that indicates that the defects have a net spin about twice what one would
expect based on free spins \cite{imai2}.
\bibitem{dho}
The authors of Ref.~\onlinecite{nilsen} fit their INS data instead with a damped harmonic oscillator form,
which leads to a specific heat in poorer agreement with the data than the Lorentzian fit employed here.
\end{thebibliography}
\end{document}